\documentclass[10pt,twocolumn]{article}
\usepackage[letterpaper,top=0.68in,bottom=0.72in,left=0.72in,right=0.72in,columnsep=0.27in]{geometry}
\usepackage{booktabs}
\usepackage{tabularx}
\usepackage{adjustbox}
\usepackage{placeins}
\usepackage{float}
\usepackage{seqsplit}
\usepackage{graphicx}
\usepackage{microtype}
\usepackage{balance}
\usepackage[hidelinks]{hyperref}
\usepackage[T1]{fontenc}
\usepackage{newtxtext}
\usepackage{newtxmath}
\usepackage[table]{xcolor}
\newcolumntype{Y}{>{\raggedright\arraybackslash}X}
\definecolor{P08Navy}{HTML}{243947}
\definecolor{P08Green}{HTML}{397650}
\definecolor{P08Red}{HTML}{8B1B18}
\hypersetup{
  pdftitle={A Reproducibility Protocol for Cross-Implementation Evaluation of Post-Quantum ACVP Test Vectors},
  pdfauthor={Christopher M. Frost},
  pdfsubject={Post-quantum cryptography and ACVP reproducibility},
  pdfkeywords={post-quantum cryptography, ACVP, reproducibility, conformance testing, interoperability}
}
\title{\vspace{-1.5em}\bfseries A Reproducibility Protocol for Cross-Implementation\\Evaluation of Post-Quantum ACVP Test Vectors}
\author{{\large Christopher M. Frost}\\\vspace{2pt}
{\normalsize HEOSSI (Pte.) Ltd., Singapore}\\\vspace{3pt}
{\small\href{mailto:christopher@heossi.com}{christopher@heossi.com}\hspace{1.5em}
\href{https://orcid.org/0009-0002-1027-1149}{ORCID 0009-0002-1027-1149}}}
\date{}
\begin{document}
\maketitle
\vspace{-1.4em}

\begin{abstract}
Independent implementations of a cryptographic standard should reproduce the same known-answer results, yet agreement is meaningful only when the corpus, revisions, public interfaces, exclusions, and evidence are precisely stated. This study defines a product-neutral reproducibility protocol for three public implementations of NIST ML-KEM against a pinned public Automated Cryptographic Validation Protocol corpus. Protocol v2 freezes provider-specific capabilities, applies one validation-error taxonomy symmetrically, preserves every selected case, and separates bytes, validation verdicts, unsupported operations, and adapter errors. The source-built experiment evaluated \texttt{\mbox{@noble}/\allowbreak{}\mbox{post-quantum}} 0.7.0, liboqs 0.16.0, and Go 1.26.4. Across three repetitions, the required Cartesian product comprised 2,160 base records: all 1,650 declared executable evaluations matched the NIST oracle, and all 510 unsupported records matched Go's predeclared capability boundary. Pairwise agreement was complete on every executable overlap: 720 of 720 noble-liboqs records and 210 of 210 records for each Go pairing. A separate \mbox{\texttt{keyGen-ek-projection}} diagnostic matched all 150 Go encapsulation-key projections without counting them as full key generation. Three frozen controls independently exercised byte comparison, verdict comparison, and malformed-response error separation; each produced its exact predeclared outcome. No base failure, adapter error, or status instability occurred. The evidence establishes bounded author-run repeatability and exposes a practical standards gap: public ML-KEM packages provide materially different deterministic and validation-test surfaces. Independent external reproduction remains unobserved. The results do not establish certification, exhaustive correctness, side-channel resistance, secure integration, or production assurance.

\end{abstract}
\section{Introduction}
NIST FIPS 203 specifies ML-KEM as a post-quantum key-encapsulation mechanism and defines three parameter sets: ML-KEM-512, ML-KEM-768, and ML-KEM-1024 \cite{ref1}. A standard, however, is not itself evidence that a particular implementation produces the required outputs. Conformance testing narrows that gap by applying defined inputs and comparing implementation outputs with expected results. The NIST Cryptographic Algorithm Validation Program (CAVP) uses ACVP to generate and validate such tests \cite{ref2}, while the public ACVP-Server repository provides sample vector files that can support research and implementation testing \cite{ref3}.

Known-answer evidence is easy to overstate. False assurance can arise when a harness omits unsupported cases, collapses execution errors into negative verdicts, compares favourable fields, or changes denominators across providers. A fixed vector suite covers only the operations and cases executed. NIST's ML-KEM ACVP specification identifies requirements that an ACVP server does not test, including zeroization and random-bit-generator strength \cite{ref4}. Reproducibility therefore requires a falsifiable evidence contract as well as an executable harness.

This paper studies whether three separately packaged public implementations reproduce the expected outputs and validation verdicts in a pinned public ML-KEM ACVP corpus. \texttt{\mbox{@noble}/\allowbreak{}\mbox{post-quantum}} is a TypeScript and JavaScript implementation with deterministic ML-KEM entry points \cite{ref5}. liboqs is a C library maintained by the Open Quantum Safe project and dispatches its selected ML-KEM release to a vendored mlkem-native backend \cite{ref6}, \cite{ref7}. Go 1.26.4 exposes FIPS 203 ML-KEM-768 and ML-KEM-1024 through its standard library \cite{ref24}. Their languages, packaging, key representations, supported parameter sets, and public validation signals differ. The study consequently claims observable cross-package agreement, not complete genealogical independence.

The study makes five contributions:

\begin{enumerate}
\item It defines a bounded, public-only experiment with immutable source revisions, SHA-256 input digests, byte-level and verdict equivalence rules, explicit unsupported-case handling, and a predeclared repetition policy.
\item It provides a new product-neutral harness whose adapters use only the selected projects' public releases and contain no private product, binding, service, CI, deployment, or production code.
\item It derives provider-specific executable and unsupported sets from a symmetric public-API audit, including an explicit taxonomy that distinguishes a documented validation rejection from an invocation error.
\item It reports case-preserving results for all 2,160 base evaluations and isolates a component-only key-generation projection in a separate diagnostic namespace.
\item It defines falsifiable audit predicates and frozen byte, verdict, and error-taxonomy controls, while distinguishing repeatable public-corpus evidence from CAVP validation, FIPS 140-3 module validation, and independent replication.
\end{enumerate}
\subsection{Methodological novelty and claim boundary}
The cryptographic primitive, NIST vectors, and evaluated implementations are public prior work. The paper does not claim novelty in ML-KEM, ACVP, noble, or liboqs. Its methodological contribution is a protocol for producing reviewable cross-implementation evidence when public interfaces expose only a partial common operation surface. Five linked disciplines distinguish the protocol from a successful-vector demonstration: immutable input identity, an explicit executable-surface contract, preservation of unsupported cases in the denominator, case-level terminal evidence, and a normalization rule that makes a later reproduction mechanically comparable without treating timestamps or diagnostic duration as scientific outcomes.

The contribution is falsifiable. A reviewer can refute the result by identifying an omitted case, digest mismatch, misdeclared adapter operation, oracle disagreement, unstable status, or normalized reproduction set that differs from the preserved digest. A matching rerun supports only the bounded protocol result, not untested security properties.

\begin{table*}[!t]
\centering
\footnotesize
\caption{Claim-to-evidence contract for the methodological contribution.}
\label{tab:1}
\begin{tabularx}{\linewidth}{@{}YYY@{}}
\toprule
Claim component & Required evidence & Falsification condition \\
\midrule
Corpus identity & NIST commit plus per-file SHA-256 digests & Retrieved bytes differ or a selected file is undeclared \\
Executable surface & Provider metadata and operation-specific adapter path & Adapter cannot exercise the stated public operation \\
Complete denominator & One terminal record for every provider, repetition, and selected case & Missing, duplicated, or silently omitted tuple \\
Known-answer agreement & Operation-specific byte comparisons against the NIST oracle & Any required returned field differs \\
Reproduction agreement & Canonical sorted terminal tuples and normalized digest & Any terminal status, reason, or case identity differs \\
\bottomrule
\end{tabularx}
\end{table*}
The result is deliberately narrow. Noble and liboqs executed all 240 selected identities per repetition. Go executed 70: 50 encapsulation cases and 20 encapsulation-key checks at ML-KEM-768 and ML-KEM-1024. Go's remaining 170 identities stayed visible as unsupported because its public API lacks ML-KEM-512, expanded decapsulation-key import, deterministic full key-generation output in the NIST expanded-key representation, and decapsulation-key checking for the supplied corpus form. Unsupported therefore means absent from the frozen public test surface, not algorithm failure.

\section{Background}
\subsection{ML-KEM}
ML-KEM is derived from CRYSTALS-Kyber, selected through the NIST post-quantum cryptography standardization process \cite{ref8}, \cite{ref9}. FIPS 203 defines key generation, encapsulation, and decapsulation together with required input checks and three parameter sets \cite{ref1}. The parameter sets trade performance and size against security strength, but the present study makes no performance or comparative-security claim. It treats the three parameter sets as separate conformance strata.

ML-KEM's deterministic internal algorithms are particularly suitable for known-answer testing. Key generation accepts seeds \mbox{\texttt{d}} and \mbox{\texttt{z}}; internal encapsulation accepts an encapsulation key and message \mbox{\texttt{m}}; and internal decapsulation accepts a decapsulation key and ciphertext. A test system can therefore supply fixed inputs and compare the resulting keys, ciphertext, and shared secret with fixed expected bytes. The separation of key generation, encapsulation, and decapsulation is also reflected in broader KEM-based protocol abstractions such as Hybrid Public Key Encryption \cite{ref22}. FIPS 203 additionally specifies encapsulation-key and decapsulation-key checks. A cross-implementation protocol must define which public acceptance and rejection signals implement those checks, because method names, key representations, exception types, and return codes are not uniform.

\subsection{ACVP and CAVP}
ACVP separates algorithm-specific JSON schemas from the protocol used to negotiate capabilities, exchange vector sets, and return results \cite{ref4}, \cite{ref10}. For ML-KEM, the public specification covers key-generation Algorithm Functional Tests (AFT), encapsulation AFT cases, decapsulation Validation Tests (VAL), and key-check VAL cases \cite{ref4}. The public sample files used here consist of prompt and expected-result documents. Test-group identifiers and test-case identifiers join the inputs to their expected outputs.

CAVP validation has an institutional meaning that is not created by running public sample files. NIST provides the ACVTS Demo environment as a no-cost, credentialed sandbox with dynamically generated tests, while production access through accredited laboratories is the route to algorithm certificates \cite{ref2}, \cite{ref25}. Cryptographic-module validation additionally evaluates a defined module boundary under FIPS 140-3 requirements, which is distinct from replaying public algorithm vectors \cite{ref16}. This study used neither an ACVTS Demo nor production session and sought no algorithm or module certificate. The term \mbox{\texttt{pass}} below means byte equality or validation-verdict equality with the selected expected result under the frozen protocol.

\subsection{Reproducibility terminology}
The ACM artifact policy treats software, scripts, input data, and raw outputs as research artifacts and distinguishes artifact availability from independent reproduction of results \cite{ref11}. The National Academies similarly separates reproducibility, using the same inputs and methods, from replicability through new evidence directed at the same question \cite{ref18}. FAIR principles and software-citation principles further motivate persistent identification, explicit provenance, and machine-actionable metadata for the released artifact \cite{ref19}, \cite{ref20}. This paper uses ``reproducible'' operationally: an independent researcher can obtain the pinned public sources, build the three releases, execute the disclosed harness, and evaluate whether the same case statuses result. The present author-run measurement is not an independent replication and does not claim an artifact badge.

\subsection{ACVP vector semantics and replay}
An ACVP sample corpus is structured JSON evidence rather than a flat list of hexadecimal strings; JSON's interoperable syntax does not itself supply the experiment-specific semantics \cite{ref21}. The prompt files group tests by parameter set and operation, while the corresponding expected-result files supply the values against which a response is judged. Test-group identifiers (\mbox{\texttt{tgId}}) and test-case identifiers (\mbox{\texttt{tcId}}) provide the stable join keys. The test type supplies part of the semantics: algorithm functional tests (AFT) exercise ordinary operation with prescribed inputs, whereas validation tests (VAL) ask an implementation to process or assess a supplied object. The meaning of a field therefore depends on its group context, not only on its name \cite{ref3}, \cite{ref4}.

For ML-KEM key generation, deterministic replay begins from the prescribed seed material and succeeds only when both the encapsulation key and decapsulation key equal the expected bytes. For encapsulation, the supplied encapsulation key and deterministic randomness lead to a ciphertext and shared secret, both of which must match. For decapsulation, the supplied decapsulation key and ciphertext lead to one expected shared secret. These are compound predicates. Reporting a case as passed after checking only one returned field would weaken the oracle and could conceal a defect in another output.

The key-check groups create a different interface problem. They ask whether supplied keys satisfy validation requirements. Protocol v2 maps a key check only when a documented public operation either accepts the encoded key normally or emits a predeclared validation-specific rejection signal. Every other exception, return code, crash, panic, timeout, resource failure, or malformed adapter response is an \mbox{\texttt{error}}, never a false verdict. This rule is applied symmetrically. Noble's public operations expose validation-specific exceptions; liboqs exposes documented backend return codes through its public KEM operations; and Go exposes constructors that validate encoded encapsulation keys. A case is never silently removed, and missing public representation support remains \mbox{\texttt{unsupported}}.

Replay also requires a distinction between corpus identity and corpus location. A repository path or branch name can identify where bytes were obtained, but it does not make those bytes immutable. The experiment binds the repository commit and the SHA-256 digest of every selected prompt and expected-result file. If a URL later serves different content, preparation fails before execution. The same principle applies to implementation archives. Human-readable version labels support interpretation; immutable commits and archive digests bind the actual input.

The result remains sample-corpus evidence. It excludes server-side negotiation, laboratory controls, module-boundary review, entropy-source assessment, and certificate issuance. Public replay is a transparent known-answer experiment, not institutional validation.

\section{Related Work}
The CRYSTALS-Kyber specification and implementation packages supplied algorithm descriptions, reference code, optimized code, and test vectors before standardization \cite{ref8}, \cite{ref9}. FIPS 203 subsequently fixed ML-KEM's normative algorithms and parameter sets \cite{ref1}. NIST IR 8413 documents the selection context and the transition from the third-round candidate to standardization \cite{ref12}. This lineage motivates exact revision pinning: a result for a Kyber candidate revision is not automatically a result for final FIPS 203 ML-KEM.

Open Quantum Safe develops liboqs as a common C library for post-quantum algorithms. Stebila and Mosca describe the project's role in early Internet-software integration \cite{ref7}. The pinned liboqs release documents ML-KEM as a stable standardized name and, for the selected build, dispatches the algorithm to its vendored mlkem-native backend \cite{ref6}. \texttt{\mbox{@noble}/\allowbreak{}\mbox{post-quantum}} provides a compact TypeScript implementation and documents seeded key generation and deterministic encapsulation inputs \cite{ref5}. Go 1.26.4 supplies ML-KEM-768 and ML-KEM-1024 in \texttt{\mbox{crypto}/\allowbreak{}\mbox{mlkem}}, with testing-only derandomized encapsulation in the public \texttt{\mbox{crypto}/\allowbreak{}\mbox{mlkem}/\allowbreak{}\mbox{mlkemtest}} package \cite{ref24}. The present work evaluates the public operation surfaces of these exact releases, not their broader algorithm coverage, security engineering, or performance.

Differential testing compares multiple systems on common inputs and treats disagreement, crashes, or nontermination as candidates for investigation \cite{ref13}. Earlier cryptographic differential work includes CDF's hand-crafted cross-library comparisons \cite{ref28} and Cryptofuzz's broader multi-backend, mutation-driven approach \cite{ref26}. Recent work extended Cryptofuzz for ML-KEM and ML-DSA and reported practical representation and validation-interface differences among Botan, OpenSSL, and liboqs \cite{ref27}. That work is a master's thesis rather than peer-reviewed archival evidence, but its independently observed interface friction corroborates the present testability finding. Cryptographic test projects such as Wycheproof complement standards-based tests with adversarial and edge-case inputs \cite{ref14}. This study is closer to constrained differential testing than to fuzzing: inputs and expected outputs are fixed by a public NIST corpus, and agreement is evaluated both against the oracle and across providers. The oracle prevents a false conclusion in which multiple implementations agree with one another but are all wrong on a case. No priority claim is made for differential testing of post-quantum implementations.

Three evidence traditions intersect here. Known-answer testing supplies a public oracle; differential testing supplies cross-provider comparison and visible disagreement; artifact evaluation demands pinned inputs, executable instructions, raw records, and bounded rerun claims \cite{ref11}, \cite{ref13}. Alone, oracle testing misses interface comparability, differential testing can mistake shared wrong output for correctness, and artifact packaging can make an inadequately bounded experiment repeatable.

The comparison also differs from compatibility demonstrations that execute a single successful exchange. A successful end-to-end exchange can show that two endpoints interoperate on one path, but it may not reveal intermediate byte differences, unsupported operations, or omitted cases. Here, every selected case has an identity and a terminal status. This case-preserving design makes denominator changes detectable and lets a reviewer distinguish algorithm agreement from interface coverage.

Reproducibility literature often distinguishes repeatability by the original team, reproducibility by a different team using the same artifacts, and replication using independently developed artifacts. Terminology varies across communities, which is why this paper states its operational meaning rather than relying on the label alone. The completed Protocol v2 measurement is an author-run, source-built execution of a frozen protocol. It supports artifact completeness without constituting independent external reproduction. An independently operated rerun or separately written adapter would reduce shared harness assumptions.

The closest practical comparison is therefore neither a performance contest nor a certification exercise. NIST's ACVTS Demo offers live, dynamically generated sandbox sessions \cite{ref25}; this work instead studies deterministic replay from immutable public sample files so that every input and expected result can be archived and reviewed without credentials. Its contribution lies in the conjunction of exact source identity, provider-specific interface boundaries, byte and verdict oracle comparison, denominator preservation, and restrained assurance claims.

This is not benchmarking, formal verification, or certification. Adapter timing is diagnostic; no proof relates implementation source to FIPS 203; and public vector replay does not reproduce CAVP or CMVP laboratory, operational-environment, or certificate processes \cite{ref2}.

\section{Threat Model and Assurance Model}
The experiment addresses accidental or systematic disagreement between public implementations and the pinned NIST oracle. Relevant failure modes include an incorrect deterministic algorithm path, byte-encoding disagreement, parameter-set confusion, omitted cases, adapter crashes, and unstable execution. The harness counters omission by enumerating every joined NIST case, counters silent adapter failure with explicit error records, and counters accidental nondeterminism with three repetitions.

The experiment does not model a malicious implementation, compromised upstream release, compiler back door, build-host compromise, or malicious vector source. Source and input digests make post-selection changes detectable but do not establish the trustworthiness of the original bytes. The three providers and the NIST oracle are not assumed to be statistically independent. A shared defect, shared upstream component, or shared misunderstanding could produce agreement.

The protected assets in this study are the integrity and interpretability of the reported result, not production keys or user data. All vector inputs and expected outputs are public. The assurance objective is therefore limited: accurately report whether the pinned releases reproduce the pinned expected bytes through the selected public operations, while preserving failures and unsupported cases. Side-channel leakage, memory erasure, and randomness quality in ordinary randomized operation are outside the model; in particular, this deterministic vector replay does not evaluate a deterministic random bit generator against the mechanisms specified by NIST SP 800-90A Rev. 1 \cite{ref23}. Dependency compromise, application integration, and operational security are also outside the model.

\section{Research Questions and Predeclared Expectations}
The study asks:

\begin{itemize}
\item RQ1: Does each selected release reproduce the expected bytes or validation verdicts for every identity in its predeclared executable set?
\item RQ2: Do provider pairs agree on every identity in the intersection of their executable sets while remaining consistent with their complete provider-specific capability declarations?
\item RQ3: Are the observed statuses stable across three repetitions in one pinned environment?
\item RQ4: Do predeclared controls independently detect a byte disagreement, a validation-verdict disagreement, and a malformed adapter response without confusing errors with verdicts?
\item RQ5: What assurance claims remain unsupported after complete agreement on the evaluated corpus?
\end{itemize}
Protocol v1 used a uniform public-interface intersection and remains preserved as historical evidence. Before Protocol v2 was frozen, all three exact releases were audited under one abstract validation mapping. The provider-specific capability matrix, error taxonomy, record namespaces, repetitions, and predicates were then digest-bound in a preregistration manifest. The expectations were exact equality on every executable byte or verdict, capability-consistent unsupported records elsewhere, no status instability, and complete agreement on pairwise executable overlaps. Three separately frozen controls subsequently required exactly one retained byte mismatch, exactly one retained verdict mismatch, and exactly one malformed-response error with zero false verdicts. No inferential hypothesis test was planned because the selected corpus was evaluated as a census rather than sampled.

\section{Method}
\subsection{Public inputs and implementation pins}
The vector source was the public \texttt{\mbox{usnistgov}/\allowbreak{}\mbox{ACVP-Server}} repository at commit \texttt{\seqsplit{a7f283cdc87d2d6dd93c1bac59e5622c5f9f8324}}, dated 31 July 2026 \cite{ref3}. Four files were used: the prompt and expected-result files for \mbox{\texttt{ML-KEM-keyGen-FIPS203}} and \mbox{\texttt{ML-KEM-encapDecap-FIPS203}}. The harness verifies each file against its recorded SHA-256 digest before execution.

Table 2 identifies the implementation releases and public adapter surfaces.

\begin{table*}[!t]
\centering
\footnotesize
\caption{Public implementation releases and adapter surfaces.}
\label{tab:2}
\begin{tabularx}{\linewidth}{@{}YYYY@{}}
\toprule
Provider label & Public release & Immutable source commit & Adapter surface \\
\midrule
noble & \texttt{\mbox{@noble}/\allowbreak{}\mbox{post-quantum}} 0.7.0 & \texttt{\seqsplit{75c18a6bcd0d8e47dda17ecfa272c8fe417e6df4}} & documented ML-KEM JavaScript methods \\
liboqs & liboqs 0.16.0 & \texttt{\seqsplit{5a1a854b0dc9f2141bdc771c555ee60c37950183}} & public C KEM operations, pinned mlkem-native backend \\
Go & Go 1.26.4 & source SHA-256 \mbox{\texttt{4f668a32...a42602d}} & public \texttt{\mbox{crypto}/\allowbreak{}\mbox{mlkem}} and \texttt{\mbox{crypto}/\allowbreak{}\mbox{mlkem}/\allowbreak{}\mbox{mlkemtest}} APIs \\
\bottomrule
\end{tabularx}
\end{table*}
All source archives were downloaded from public release locations and verified against recorded SHA-256 digests before extraction. Noble was installed from its lock file with lifecycle scripts disabled and compiled by its declared build command. Liboqs was configured as a minimal static build containing the three ML-KEM parameter sets. Go 1.26.4 was built from its pinned archive. Go 1.26.5 was available at submission review, but the older pin is retained because Protocol v2 froze it before execution.

Table 3 makes the Protocol v2 validation promotion independently reviewable. The abstract rule is symmetric: a documented validation-bearing public operation that returns normally yields \mbox{\texttt{true}}; only the exact listed rejection signal yields \mbox{\texttt{false}}; every other exception, return code, panic, invocation failure, or malformed response is an adapter \mbox{\texttt{error}}. Go exposes no public expanded-decapsulation-key import, so no decapsulation-key-check mapping exists.

\begin{table*}[!t]
\centering
\footnotesize
\caption{Frozen public validation operations and allowlisted rejection signals.}
\label{tab:3}
\begin{tabularx}{\linewidth}{@{}YYYY@{}}
\toprule
Provider & ACVP check & Exact public operation & Allowlisted rejection signal \\
\midrule
noble & Encapsulation key & encapsulate & exception message ``ML-KEM.encapsulate: wrong publicKey modulus'' \\
noble & Decapsulation key & decapsulate & exception message ``invalid secretKey: hash check failed'' \\
liboqs & Encapsulation key & OQS\_KEM\_encaps\_derand & return code -1; pinned backend modulus-check rejection \\
liboqs & Decapsulation key & OQS\_KEM\_decaps & return code -1; pinned backend embedded-public-key hash rejection \\
Go & Encapsulation key & NewEncapsulationKey768 / 1024 & error messages ``mlkem: invalid encapsulation key length'', ``mlkem: invalid encoding length'', or ``mlkem: invalid polynomial encoding'' \\
Go & Decapsulation key & None in the pinned public API & unsupported: expanded key not importable \\
\bottomrule
\end{tabularx}
\end{table*}
The ACVP key-check records supply a key but not every argument required by these public operations. The frozen adapters therefore provide auxiliary inputs that are not interpreted as test vectors: noble receives a 32-byte zero message for encapsulation-key checks and a parameter-set-sized zero ciphertext for decapsulation-key checks; liboqs receives an all-zero seed of \mbox{\texttt{length\_encaps\_seed}} for deterministic encapsulation and a parameter-set-sized zero ciphertext for decapsulation. The validation taxonomy names these fixed inputs and is itself digest-bound by the preregistration manifest. For a rejected key, the mapped validation occurs before the auxiliary value can determine an accepted cryptographic result. For an accepted key, the protocol observes only normal completion and never compares the resulting ciphertext or shared secret. The auxiliary inputs therefore make the public validation-bearing operation invocable; they do not create expected key-check semantics or enter the byte-agreement result.

Liboqs exposes a materially coarser signal than noble or Go. At the public \mbox{\texttt{OQS\_KEM}} boundary, \mbox{\texttt{OQS\_ERROR}} is the generic integer value \mbox{\texttt{-1}}; the adapter cannot infer validation causality from that integer alone. The mapping is consequently restricted to correct-length, successfully decoded inputs and the pinned mlkem-native execution path whose documented return behaviour associates the observed \mbox{\texttt{-1}} with the modulus or embedded-public-key hash rejection. Allocation failure in the small public-API driver exits separately before the KEM call, and unexpected driver outcomes become adapter errors, but an unanticipated backend failure collapsed to \mbox{\texttt{OQS\_ERROR}} would remain indistinguishable from validation rejection. This is a residual limitation of the liboqs public signal, not evidence of equivalent error specificity across providers.

\subsection{Harness independence}
The experiment harness was written anew for this study. A provider-neutral Python runner joins NIST prompt and expected-result records by \mbox{\texttt{tgId}} and \mbox{\texttt{tcId}}, invokes a provider adapter, records status, and continues after any case failure. The noble adapter imports the compiled public module. The liboqs adapter invokes a small public-API driver linked to the pinned build. The Go adapter is compiled by the source-built Go toolchain. No adapter imports private product code or calls a private binding.

Adapters publish metadata containing their provider label, release version, and supported operations. The frozen capability matrix, rather than runtime success, determines whether each provider-case identity is executable. Unsupported operations receive a record with a stable reason. The controller also enforces namespaces: \mbox{\texttt{base}} records determine the reported denominator; \mbox{\texttt{keyGen-ek-projection}} diagnostics cannot count as key-generation passes; mutation controls and future specification-divergence challenges are likewise isolated.

\begin{figure*}[!t]
\centering
\includegraphics[width=0.82\textwidth]{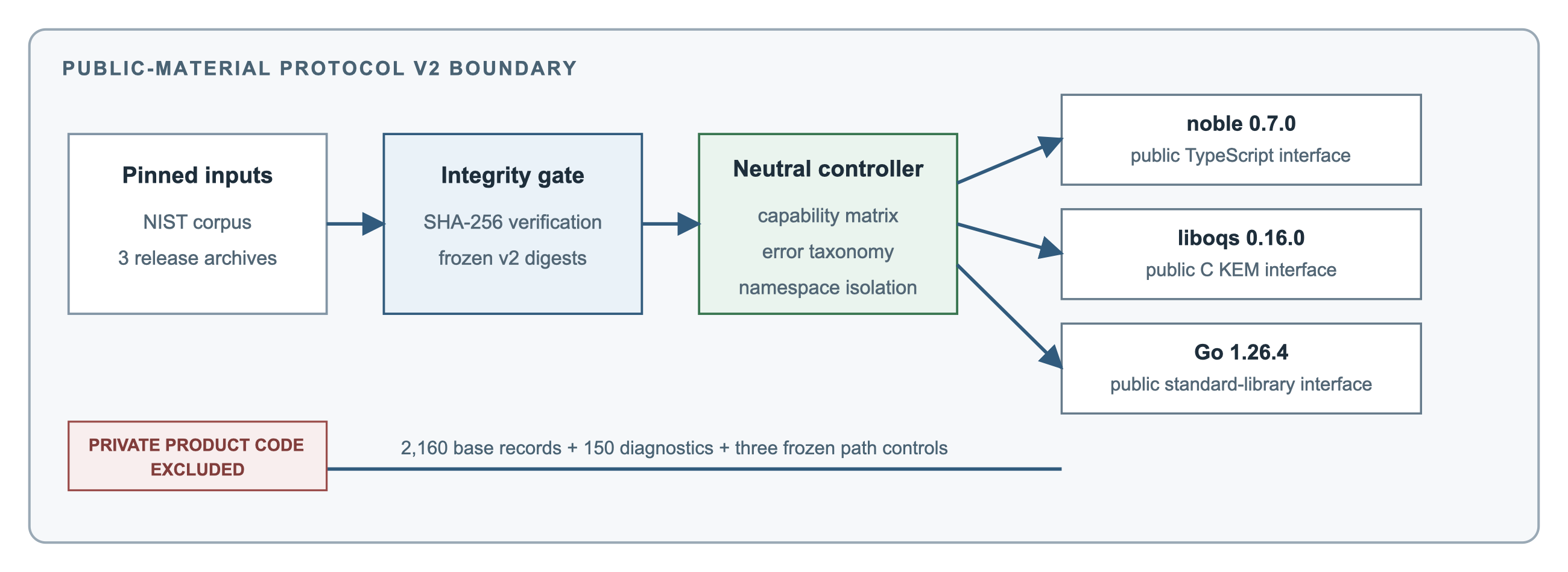}
\caption{Public-material experimental architecture. Immutable public sources are digest-verified before preparation. A provider-neutral controller applies the frozen capability matrix to three public interfaces, compares returned bytes or validation verdicts with the NIST oracle, and writes case-preserving evidence. Private product code remains outside the experiment.}
\label{fig:experiment-pipeline}
\end{figure*}
\subsection{Corpus composition}
Table 4 reports the selected corpus and provider-specific execution disposition.

\begin{table*}[!t]
\centering
\footnotesize
\caption{Selected NIST corpus and Protocol v2 execution disposition.}
\label{tab:4}
\begin{tabularx}{\linewidth}{@{}YYYYYY@{}}
\toprule
Operation & Cases & Noble & liboqs & Go & Expected comparison \\
\midrule
Key generation AFT & 75 & 75 & 75 & 0 & \mbox{\texttt{ek}} and expanded \mbox{\texttt{dk}} \\
Encapsulation AFT & 75 & 75 & 75 & 50 & \mbox{\texttt{c}} and \mbox{\texttt{k}} \\
Decapsulation VAL & 30 & 30 & 30 & 0 & \mbox{\texttt{k}} \\
Encapsulation-key check VAL & 30 & 30 & 30 & 20 & observed verdict \\
Decapsulation-key check VAL & 30 & 30 & 30 & 0 & observed verdict \\
Total per repetition & 240 & 240 & 240 & 70 & provider-specific \\
\bottomrule
\end{tabularx}
\end{table*}
Noble and liboqs execute all 240 identities. Go exposes only the 768 and 1024 parameter sets. Its public encapsulation interface executes 50 AFT cases and its encoded encapsulation-key constructors execute 20 validation cases. The NIST key-generation oracle requires expanded decapsulation-key bytes, whereas Go publicly emits a 64-byte \mbox{\texttt{d || z}} seed form; the adapter does not reimplement expansion. The decapsulation corpus supplies expanded keys that Go cannot import publicly. Across three providers and three repetitions, the base design therefore requires 2,160 records: 1,650 executable and 510 unsupported.

\begin{figure*}[!t]
\centering
\includegraphics[width=0.82\textwidth]{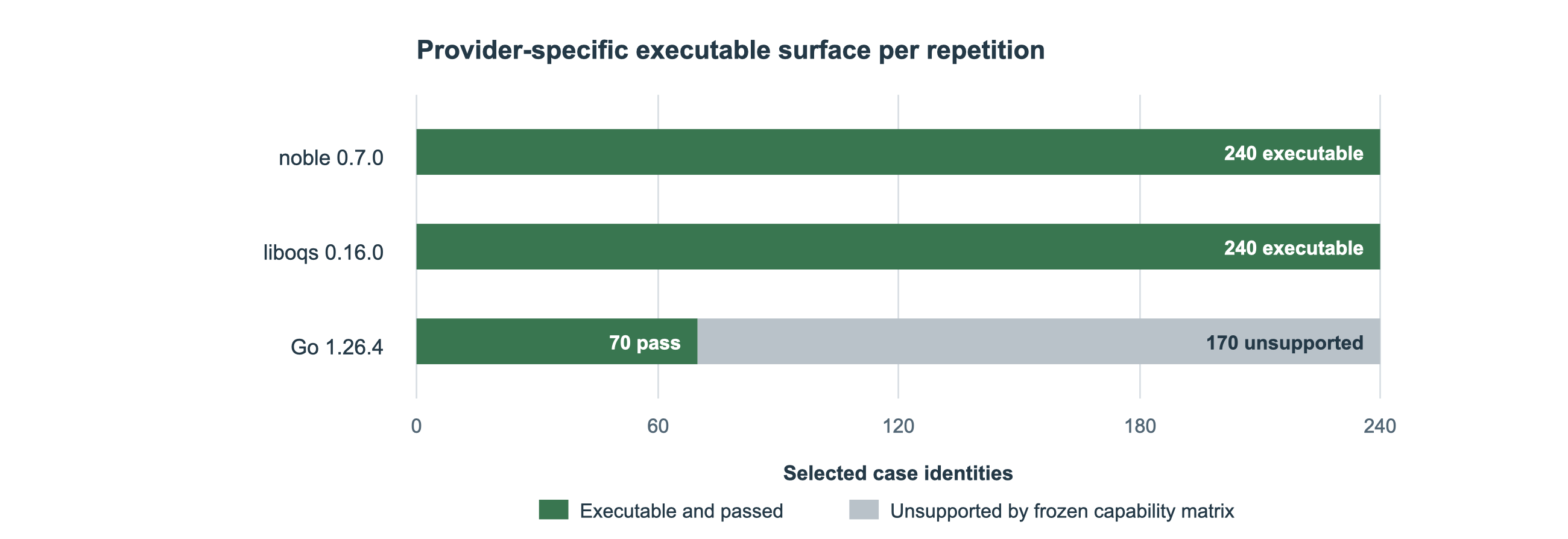}
\caption{Provider-specific executable surface. Noble and liboqs execute all 240 identities per repetition; Go executes 70 and retains 170 as unsupported. The full 240-identity denominator is preserved for every provider.}
\label{fig:case-surface}
\end{figure*}
\subsection{Equivalence and failure rules}
Hexadecimal strings were decoded to bytes by the adapters. Equality required every output byte to match the corresponding expected value; hexadecimal letter case was ignored and byte order was not changed. A key-generation case passed only if both keys matched. An encapsulation case passed only if both ciphertext and shared secret matched. A decapsulation case passed only if the shared secret matched.

Any unequal field produced \mbox{\texttt{fail}} and retained the mismatched field name. For validation cases, a normal documented return maps to verdict \mbox{\texttt{true}}; only an allowlisted validation-specific rejection in Table 3 maps to \mbox{\texttt{false}}. Every unexpected exception, return code, crash, panic, timeout, allocation failure, malformed response, or invocation failure maps to \mbox{\texttt{error}}. Unsupported records use the provider-specific frozen reason. Adapter diagnostic text is represented by digest rather than copied verbatim. The runner continues after every status so that denominators cannot shrink after a failure.

\subsection{Repetitions and environment}
Every executable case was run three times. Repetition was used to detect status instability, not to estimate random performance variation. The final run occurred on macOS 26.6, arm64, with Python 3.14.6 and Node.js 24.18.0. Liboqs was compiled by AppleClang 21.0.0 through CMake and Ninja. Go 1.26.4 and its adapter were compiled from the verified source archive. Build diagnostics are recorded as environment evidence but are not interpreted as security findings.

\subsection{Evidence preservation}
The raw artifact is newline-delimited JSON with one record for each provider, repetition, and NIST case. Separate files record the environment, source commits, release versions, input URLs, input digests, adapter metadata, aggregate results, and SHA-256 digests of the result artifacts. SHA-256 supplies the standardized digest function used to bind file identity, not a proof that the bound content is semantically correct \cite{ref17}. The raw records omit the large public input and output byte strings because those remain reconstructible from the pinned NIST files and implementation execution. Case identity and status are retained.

\subsection{Evidence model and proof obligations}
The experiment separates outputs from claims. A returned key, ciphertext, or shared secret is an implementation output. A \mbox{\texttt{pass}} record claims that all required returned fields match the NIST expected fields. Aggregate counts derive from case records; cross-provider agreement and stability derive from joins across providers and repetitions. A plausible aggregate alone does not prove that every case exists or every field was compared.

Let \mbox{\texttt{C}} be the 240 selected case identities, \texttt{\mbox{P = \{noble},\allowbreak{}\mbox{ liboqs},\allowbreak{}\mbox{ go\}}}, and \texttt{\mbox{R = \{1},\allowbreak{}\mbox{ 2},\allowbreak{}\mbox{ 3\}}}. For every provider \mbox{\texttt{p}}, the frozen capability matrix defines disjoint sets \mbox{\texttt{E\_p}} and \mbox{\texttt{U\_p}} whose union is \mbox{\texttt{C}}. Their cardinalities are \texttt{\mbox{(240},\allowbreak{}\mbox{ 0)}} for noble, \texttt{\mbox{(240},\allowbreak{}\mbox{ 0)}} for liboqs, and \texttt{\mbox{(70},\allowbreak{}\mbox{ 170)}} for Go. Protocol v2 creates the following audit predicates:

\begin{enumerate}
\item \textbf{Cartesian completeness:} exactly one base record exists for every tuple in \mbox{\texttt{P x R x C}}, yielding \texttt{\mbox{3 x 3 x 240 = 2},\allowbreak{}\mbox{160}} records.
\item \textbf{Capability consistency:} an identity in \mbox{\texttt{E\_p}} terminates as \mbox{\texttt{pass}}, \mbox{\texttt{fail}}, or \mbox{\texttt{error}}; an identity in \mbox{\texttt{U\_p}} terminates as \mbox{\texttt{unsupported}} with its frozen provider reason.
\item \textbf{Executable cardinality:} the base contains exactly \texttt{\mbox{3 x (240 + 240 + 70) = 1},\allowbreak{}\mbox{650}} executable records.
\item \textbf{Unsupported cardinality:} the base contains exactly \mbox{\texttt{3 x 170 = 510}} unsupported records, all from Go identities in \mbox{\texttt{U\_go}}.
\item \textbf{Oracle agreement:} every passing byte-producing record satisfies all operation-specific byte predicates, and every passing validation record has an observed verdict equal to NIST \mbox{\texttt{testPassed}}.
\item \textbf{Failure preservation:} any byte or verdict disagreement remains \mbox{\texttt{fail}} with its mismatch evidence; no failed identity is omitted or relabelled unsupported.
\item \textbf{Error separation:} only an allowlisted validation rejection becomes verdict \mbox{\texttt{false}}; every other provider or adapter failure becomes \mbox{\texttt{error}}.
\item \textbf{Pairwise-overlap agreement:} agreement is evaluated on \mbox{\texttt{E\_p}} intersect \mbox{\texttt{E\_q}}, while the full provider-specific denominator remains visible. Per repetition, overlaps are 240 for noble-liboqs and 70 for each Go pairing.
\item \textbf{Repetition stability:} the case identity, status, reason, observed verdict, and mismatch fields are invariant across the three repetitions.
\item \textbf{Namespace isolation:} diagnostics, mutation controls, and specification-divergence records cannot enter base aggregates or satisfy base predicates.
\end{enumerate}
The separate \mbox{\texttt{keyGen-ek-projection}} namespace contains exactly 50 Go records per repetition. It checks the public encapsulation-key component derived from \mbox{\texttt{d}} but cannot satisfy the full NIST key-generation predicate, which also requires the expanded decapsulation key. This structural separation prevents a partial comparison from inflating the base pass count.

\textbf{Proposition 1 (bounded audit soundness).} Assume that the pinned corpus join identifies the selected set \mbox{\texttt{C}}, each adapter faithfully invokes the public operation declared in its frozen capability metadata, and the preserved records and manifests have not been altered. If predicates 1-10 hold, then: (i) no selected provider-repetition-case tuple is omitted or duplicated; (ii) every tuple outside a provider's executable set is visible as unsupported with the predeclared reason; (iii) every tuple reported as passed satisfies every byte or verdict equality required for its operation; (iv) every disagreement or non-allowlisted execution failure remains visible as \mbox{\texttt{fail}} or \mbox{\texttt{error}}; and (v) pairwise agreement is evaluated only over declared executable intersections without changing any provider's full denominator.

\textbf{Proof.} Predicate 1 gives existence and uniqueness over \mbox{\texttt{P x R x C}}. Predicates 2-4 partition records by the frozen \mbox{\texttt{E\_p}} and \mbox{\texttt{U\_p}} sets and fix their cardinalities. Predicate 5 expands \mbox{\texttt{pass}} into the required field or verdict equalities. Predicates 6 and 7 exclude disagreement and unexpected execution failures. Predicate 8 defines pairwise joins without removing non-overlapping base identities. Predicates 9 and 10 prevent repetition drift and namespace contamination. The conclusions follow under the stated assumptions. The truth of those assumptions, the completeness of the oracle, and the security of ML-KEM are not proved.

These predicates are dataset obligations, not proofs of ML-KEM or its implementations. Reviewers can recalculate them from \mbox{\texttt{cases.jsonl}}; the manifest binds the case, summary, and input evidence. Cardinality detects omissions, while byte comparison establishes equality for executed pairs. Neither covers all inputs, constant-time execution, or adversarial resistance.

\subsection{Adapter contract and semantic normalization}
Each adapter implements a line-oriented request-response contract identifying the operation, parameter set, case, provider, version, status, and returned fields or stable error class. The controller owns enumeration and comparison; adapters only translate requests to documented public surfaces. Central ownership prevents providers from selecting different cases or pass criteria while producing superficially similar totals.

Normalization decodes hexadecimal text to bytes but never reverses byte order, pads, truncates, regenerates missing test-vector inputs, or accepts alternative encodings. The fixed auxiliary invocation values in Section 6.1 are declared constants, not regenerated vectors. An adapter unable to map a public interface to the specified operation reports \mbox{\texttt{unsupported}} or \mbox{\texttt{error}} rather than repairing the case.

Noble invokes documented JavaScript methods, liboqs uses public C KEM operations linked to the pinned backend, and Go uses standard-library APIs compiled from pinned source. Their key forms, parameter sets, and failure signals differ; the contract resolves only the selected ACVP semantics. Adapter metadata freezes supported operations before execution, so runtime success, crashes, or permissive parsing cannot redefine the test surface.

\subsection{Audit and reproduction procedure}
The experiment is designed for a staged audit rather than a single opaque command. A reviewer can perform the following checks independently:

\begin{enumerate}
\item Inspect \mbox{\texttt{pins.json}} and confirm that every remote input names a public HTTPS source, immutable revision where available, and SHA-256 digest.
\item Run preparation in an empty work directory. Preparation downloads only declared inputs, verifies each digest before extraction, builds the selected release surfaces, and writes no result claim.
\item Inspect adapter metadata and compare declared operation support with the frozen protocol.
\item Run the controller into a new output directory without modifying the preserved author-run directory.
\item Verify raw-record cardinality, uniqueness of the provider-repetition-case tuple, allowed status values, and the absence of missing selected identities.
\item Recalculate operation, parameter-set, provider, and repetition counts from the raw records rather than trusting \mbox{\texttt{summary.json}}.
\item Compare rerun statuses with the preserved run while ignoring generation timestamps and diagnostic elapsed times.
\item Verify the result-artifact manifest and then the release manifest, ensuring that no undeclared file has entered the candidate bundle.
\end{enumerate}
The stages preserve failure meaning: a digest mismatch is an input-integrity failure, a compiler failure is a build result, and a mismatched returned key is a case failure. Work-directory paths, timestamps, elapsed times, diagnostic ordering, and download timing may differ. Provider version, source commit, vector digest, case identity, terminal status, and operation-specific comparison may not.

\begin{table*}[!t]
\centering
\footnotesize
\caption{Evidence layers, audit predicates, and residual gaps.}
\label{tab:5}
\begin{tabularx}{\linewidth}{@{}YYYY@{}}
\toprule
Evidence layer & Audit predicate & Preserved basis & Residual gap \\
\midrule
Input identity & All selected archives and vectors match pinned SHA-256 values & \mbox{\texttt{pins.json}}, \mbox{\texttt{inputs.json}} & Does not establish upstream trustworthiness \\
Case completeness & 2,160 unique provider-repetition-case records exist & \mbox{\texttt{cases.jsonl}}, protocol cardinality & Depends on correct corpus joining \\
Oracle agreement & Every executable record matches required bytes or verdict & Per-case status, verdict, and mismatch policy & Finite sample corpus only \\
Interface coverage & Every status agrees with the frozen provider matrix & Capability metadata and reasons & Public surfaces remain heterogeneous; liboqs exposes a coarse generic error signal \\
Cross-provider agreement & Providers agree on every pairwise executable overlap & Joined overlap records & Shared defects remain possible \\
Stability & Each provider-case status is unchanged across three repetitions & Repetition-indexed records & One host and toolchain \\
Measurement controls & Frozen byte and verdict mutations each yield one failure; malformed output yields one error and zero false verdicts & Three control specifications, manifests, and results & Controls do not create provider divergence \\
Artifact integrity & Result and candidate files match recorded digests & Two SHA-256 manifests & Digest does not prove semantic correctness \\
\bottomrule
\end{tabularx}
\end{table*}
\section{Results}
\subsection{Aggregate outcomes}
The experiment produced all 2,160 planned base records. There were 1,650 passes, 510 unsupported records, zero base failures, and zero adapter errors. Every audit predicate in the frozen Protocol v2 specification evaluated true.

\begin{table*}[!t]
\centering
\footnotesize
\caption{Aggregate outcomes across three repetitions.}
\label{tab:6}
\begin{tabularx}{\linewidth}{@{}YYYYYY@{}}
\toprule
Provider & Pass & Unsupported & Fail & Error & Total records \\
\midrule
noble & 720 & 0 & 0 & 0 & 720 \\
liboqs & 720 & 0 & 0 & 0 & 720 \\
Go & 210 & 510 & 0 & 0 & 720 \\
Total & 1,650 & 510 & 0 & 0 & 2,160 \\
\bottomrule
\end{tabularx}
\end{table*}
The counts include three repetitions. Noble and liboqs each passed all 240 selected cases per repetition. Go passed its 70 declared executable cases and retained 170 unsupported records per repetition.

\subsection{Outcomes by operation}
Noble and liboqs each passed 225 key-generation, 225 encapsulation, 90 decapsulation, 90 encapsulation-key-check, and 90 decapsulation-key-check evaluations across three repetitions. Go passed 150 encapsulation and 60 encapsulation-key-check evaluations. Its 225 key-generation, 90 decapsulation, 75 ML-KEM-512 encapsulation, 30 ML-KEM-512 encapsulation-key-check, and 90 decapsulation-key-check records remained unsupported.

\begin{table*}[!t]
\centering
\footnotesize
\caption{Outcomes by operation across three repetitions.}
\label{tab:7}
\begin{tabularx}{\linewidth}{@{}YYYYY@{}}
\toprule
Operation & Noble pass & liboqs pass & Go pass & Go unsupported \\
\midrule
Key generation & 225 & 225 & 0 & 225 \\
Encapsulation & 225 & 225 & 150 & 75 \\
Decapsulation & 90 & 90 & 0 & 90 \\
Encapsulation-key check & 90 & 90 & 60 & 30 \\
Decapsulation-key check & 90 & 90 & 0 & 90 \\
\bottomrule
\end{tabularx}
\end{table*}
\subsection{Outcomes by parameter set}
Noble and liboqs passed 80 of 80 identities for every parameter set in every repetition. Go exposes no ML-KEM-512 and therefore retained all 80 corresponding identities per repetition as unsupported. At each of ML-KEM-768 and ML-KEM-1024, Go passed 25 encapsulation and 10 encapsulation-key-check identities while retaining 45 identities unsupported. No parameter set exhibited a failure or error.

\subsection{Cross-implementation agreement}
Pairwise agreement was evaluated only where both providers declared an identity executable. Noble and liboqs agreed on all 240 identities per repetition, producing 720 of 720 matching overlap records. Noble-Go and liboqs-Go each agreed on all 70 shared executable identities per repetition, producing 210 of 210 matching overlap records for each pair. Agreement followed oracle comparison: every shared pass also matched the NIST expected bytes or verdict.

Unsupported sets were audited separately from provider disagreements. Every Go unsupported status and reason matched the frozen capability matrix, exposing the absence of a common public surface without penalizing broader APIs.

\subsection{Repetition stability}
No base or diagnostic case changed status across the three repetitions. The instability list is empty. This supports deterministic rerun stability in the reported environment but not across operating systems, architectures, compilers, or future releases.

\begin{figure*}[!t]
\centering
\includegraphics[width=0.82\textwidth]{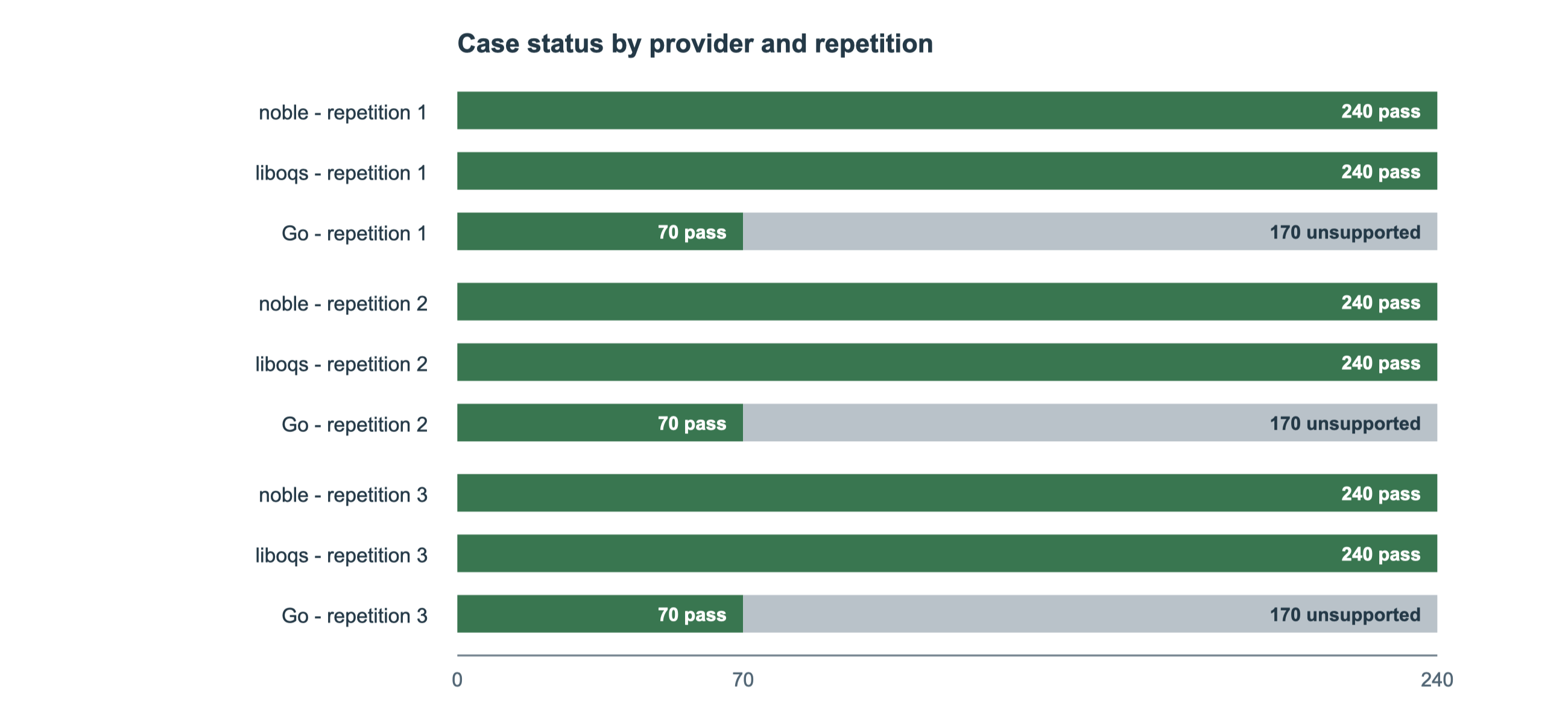}
\caption{Provider outcomes and repetition stability. Noble and liboqs each produced 240 passes per repetition. Go produced 70 passes and 170 unsupported records. The repeated bars expose the stable heterogeneous capability surface without shrinking any provider denominator.}
\label{fig:outcome-stability}
\end{figure*}
\subsection{Completeness and denominator audit}
The Cartesian product of 240 identities, three providers, and three repetitions requires 2,160 base records, and the artifact contains exactly 2,160. No provider-repetition-case tuple is duplicated or absent. The capability partition independently requires \texttt{\mbox{3 x (240 + 240 + 70) = 1},\allowbreak{}\mbox{650}} executable and \mbox{\texttt{3 x 170 = 510}} unsupported records. Observed totals equal both derivations.

Completeness prevents favourable-case attrition. The result is 1,650 passes out of 1,650 declared executable evaluations, accompanied by 510 unsupported records out of the 510 identities outside Go's frozen executable set. It is not reported as 1,650 passes out of 2,160 tests.

\subsection{Field-level and cross-provider evidence}
Byte predicates required both \mbox{\texttt{ek}} and expanded \mbox{\texttt{dk}} for key generation, both \mbox{\texttt{c}} and \mbox{\texttt{k}} for encapsulation, and \mbox{\texttt{k}} for decapsulation. Validation predicates required the observed Boolean verdict to equal NIST \mbox{\texttt{testPassed}}. Noble matched 30 of 30 encapsulation-key and 30 of 30 decapsulation-key verdicts per repetition. Liboqs matched the same complete set. Go matched all 20 available 768/1024 encapsulation-key verdicts per repetition. All 20 Go inputs had normal encoded lengths; the rejection cases therefore exercised the modulus constraint rather than a trivial length check.

The evidence does not rank providers. Diagnostic timing was not calibrated as a benchmark, so no latency, throughput, or efficiency conclusion is drawn.

\subsection{Negative evidence and retained absences}
No base \mbox{\texttt{fail}}, \mbox{\texttt{error}}, or unstable status was observed. The unsupported records describe representation and interface boundaries, not incorrect cryptographic behaviour. In particular, Go's public decapsulation-key form is a 64-byte seed, whereas the pinned NIST corpus supplies expanded 2,400-byte and 3,168-byte forms for ML-KEM-768 and ML-KEM-1024. Expanding or transforming those keys in the adapter would violate the predeclared no-semantic-invention rule.

The \mbox{\texttt{keyGen-ek-projection}} diagnostic separately tested whether Go derived the expected public encapsulation key from the supplied seed. All 150 projection records passed. Because each record carries a distinct operation label and namespace, aggregation cannot count it as a full key-generation pass. The diagnostic localizes Go's boundary to the unavailable expanded \mbox{\texttt{dk}} comparison instead of implying that key generation was unexamined.

Three controls were frozen independently of the base run and used dedicated namespaces. The byte control changed byte zero of expected ciphertext \mbox{\texttt{c}} for one exact noble case in an in-memory oracle copy; it retained 239 passes and one \mbox{\texttt{c}} failure and returned nonzero. The verdict control flipped expected \mbox{\texttt{testPassed}} for noble encapsulation-key-check identity \texttt{\mbox{(tgId=8},\allowbreak{}\mbox{ tcId=116)}}; it retained 239 passes and one \mbox{\texttt{testPassed}} failure and returned the predeclared nonzero exit. The taxonomy control fed one syntactically malformed response through the production controller's adapter-response parser; it produced exactly one \mbox{\texttt{invalid-adapter-json}} error, zero pass or fail statuses, and zero observed verdicts. The pinned corpus and base results remained unchanged. Together the controls exercise byte comparison, verdict comparison, failure preservation, and error-versus-verdict separation. They are controls of the measurement paths, not evidence of implementation defects or provider divergence.

\begin{table*}[!t]
\centering
\footnotesize
\caption{Frozen comparison-path controls and observed outcomes.}
\label{tab:8}
\begin{tabularx}{\linewidth}{@{}YYYY@{}}
\toprule
Control namespace & Exercised class & Predeclared outcome & Observed outcome \\
\midrule
byte-oracle & Byte oracle comparison & 239 pass, 1 ciphertext mismatch, 0 error & Exact match; nonzero controller exit \\
verdict-oracle & Boolean verdict comparison & 239 pass, 1 verdict mismatch, 0 error & Exact match; nonzero controller exit \\
malformed-response & Response/error taxonomy & 1 error, 0 false verdicts & Exact match; invalid JSON remained an error \\
\bottomrule
\end{tabularx}
\end{table*}
\begin{table*}[!t]
\centering
\footnotesize
\caption{Claim-to-evidence decision table.}
\label{tab:9}
\begin{tabularx}{\linewidth}{@{}YYY@{}}
\toprule
Candidate statement & Evidence status & Decision \\
\midrule
Pinned releases reproduce all required bytes or verdicts on their declared surfaces & Supported by 1,650 pass records and complete predicates & Report with corpus, revision, and capability qualifiers \\
Provider pairs agree on every executable overlap & Supported by 720, 210, and 210 joined overlap records & Report overlap denominators separately \\
Statuses remain stable across three repetitions on the author host & Supported by an empty instability set & Report with environment qualifier \\
Public ML-KEM packages expose a uniform deterministic test surface & Contradicted by the capability matrix & Report the representation and interface boundary \\
Byte and verdict failure detection, preservation, and error separation are active & Supported by three comparison-path controls & Describe as measurement controls only \\
Either provider is correct for all ML-KEM inputs & Not tested & Do not claim \\
Either provider is constant-time, certified, or production-secure & Not tested and outside method & Do not claim \\
The artifact has been independently reproduced externally & Not yet observed & Keep pending \\
\bottomrule
\end{tabularx}
\end{table*}
\subsection{Source-built execution and frozen evidence}
Preparation began from a new work directory, downloaded every declared public archive, verified its SHA-256 digest, and built all three implementation paths before measurement. The result artifact binds \mbox{\texttt{cases.jsonl}}, \mbox{\texttt{diagnostics.jsonl}}, \mbox{\texttt{environment.json}}, \mbox{\texttt{inputs.json}}, and \mbox{\texttt{summary.json}} by digest. This removes dependence on stale provider binaries but remains an author-run, same-host execution. A second independent environment has not yet reproduced Protocol v2.

\begin{figure*}[!t]
\centering
\includegraphics[width=0.82\textwidth]{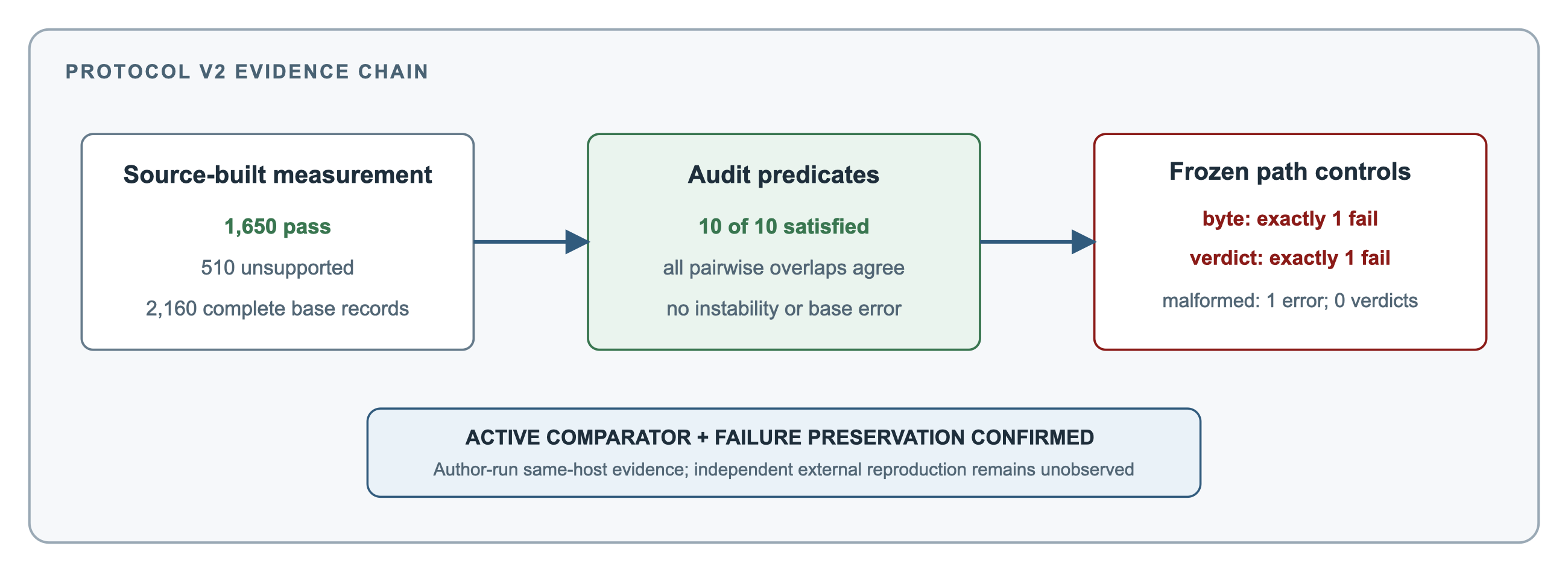}
\caption{Evidence chain and active measurement controls. Digest-verified source preparation produced the complete Protocol v2 record set. Separately frozen byte, verdict, and malformed-response controls yielded their exact required failures or error without entering base aggregates. External reproduction remains a future evidence event.}
\label{fig:clean-rerun}
\end{figure*}
\section{Discussion}
\subsection{RQ1: Reproduction of expected outcomes}
RQ1 is answered affirmatively for each declared provider surface and the pinned corpus. Noble and liboqs matched every expected byte and key-validation verdict in 240 cases per repetition. Go matched every expected result in its 70 executable cases per repetition.

The result is stronger than provider-provider agreement alone because every case was also checked against a public NIST expected result. It remains narrower than implementation correctness: a finite corpus can expose errors but cannot prove their absence on untested inputs.

\subsection{RQ2: Agreement and interface boundaries}
All provider pairs agreed on every shared executable identity, and every provider's statuses agreed with its frozen capability matrix. Protocol v2 was created because the earlier uniform-interface policy understated the public validation-bearing capabilities of noble and liboqs. Re-auditing all providers under the same abstract mapping promoted both libraries' encapsulation-key and decapsulation-key checks, while also admitting Go's encapsulation-key constructors. This correction is preserved transparently: Protocol v1 remains immutable, and Protocol v2 changes the declared surface before the expanded run.

The headline finding is not a leaderboard. Three public packages implementing the same standard expose three materially different testable fragments. Noble and liboqs cover the complete selected corpus through distinct public signals. Go lacks ML-KEM-512 and uses a seed-only public decapsulation-key representation, preventing direct comparison with the expanded-key oracle without adapter-side algorithmic reconstruction. A claim that a package supports ML-KEM therefore says less than a conformance engineer needs to know about deterministic inputs, serialization, validation, and evidence production.

\subsection{RQ3: Stability}
All base and diagnostic statuses were stable across three repetitions. Because the selected operations use fixed inputs, stability was expected. Repetition still detects state leakage, intermittent adapter errors, and accidental randomness in the harness. It is not a statistical reliability estimate and does not substitute for independently operated reruns.

\subsection{RQ4: Negative-control result}
The three controls answered RQ4 affirmatively. A predeclared byte mutation and a predeclared Boolean-verdict mutation each yielded exactly one retained failure among 240 real noble executions and the required nonzero controller result. A malformed-response fixture yielded exactly one adapter error and no pass, fail, or observed verdict. These controls rule out vacuous always-pass behaviour in the exercised byte and verdict paths and demonstrate that syntactically invalid adapter output cannot be silently converted to a negative conformance verdict. They do not challenge providers with semantically divergent cryptographic inputs and therefore do not establish sensitivity to every provider defect. A future Kyber-versus-final-ML-KEM challenge would require a separately frozen protocol identifier and record namespace.

\subsection{RQ5: Assurance boundary}
The following conclusions are supported:

\begin{itemize}
\item the pinned releases reproduced the expected bytes or verdicts for all 1,650 declared executable evaluations;
\item provider pairs agreed on every shared executable identity and each provider respected its capability matrix;
\item observed statuses were stable across three repetitions in one environment;
\item the separately frozen byte, verdict, and malformed-response controls produced their exact intended failures or error without false verdicts; and
\item the artifacts identify the exact public inputs and implementation revisions used.
\end{itemize}
The following conclusions are not supported:

\begin{itemize}
\item that either implementation is formally verified or correct for all inputs;
\item that either implementation is constant-time, resistant to side channels, free from memory-safety defects, or securely randomized in ordinary use;
\item that either implementation, build, runtime, product, or module is CAVP or FIPS validated;
\item that the three packages are independent at every source or algorithmic layer;
\item that passing the sample corpus predicts interoperability in a protocol or secure key management in an application; or
\item that a future release will retain the measured behaviour.
\end{itemize}
NIST SP 800-227 describes broader requirements for secure KEM use \cite{ref15}. Those application and lifecycle questions are outside the present experiment.

\subsection{Evidence strength and claim maturity}
The evidence forms a hierarchy: a public operation returned a result; the result matched the oracle; every executable case matched while unsupported cases remained visible; and provider pairs agreed across three repetitions. The controls further showed that wrong expected bytes and verdicts remain failures and malformed adapter output remains an error. Each level closes a distinct reporting failure mode.

The hierarchy stops before ecosystem assurance. Cross-provider agreement reduces the plausibility of some package-specific mistakes but does not quantify shared-defect probability. All providers follow the same standard and oracle and may share upstream reasoning or code ancestry.

The result supports corpus reproduction, not procurement, deployment, migration, or product-security conclusions.

\subsection{Interface and standards implications}
The capability matrix reveals an interoperability issue at the evidence interface. Standards can define an operation without ensuring that libraries expose deterministic inputs, equivalent key encodings, or comparable validation signals. A public operation can legitimately accept seed-form secret material while a conformance corpus expects expanded bytes. Both may follow their interface contracts, yet direct reproducibility is blocked unless the harness reimplements a normative transformation. Protocol v2 refuses that semantic invention and records the boundary instead.

The validation mapping also shows why error taxonomy matters. A documented invalid-key signal is a scientific verdict. A crash, allocation failure, unexpected exception, or unrelated return code is an execution error. Mapping every error to \mbox{\texttt{testFailed}} could let a broken adapter appear to match a negative test. Predeclaring exact provider signals and treating everything else as \mbox{\texttt{error}} makes accidental passes visible.

The observation has broader relevance to cryptographic agility. Algorithm support lists often indicate that a primitive exists, but they do not reveal whether validation, serialization, deterministic testing, error reporting, or provider interchange is available. Evidence-backed capability claims should therefore name operations and interfaces, not only algorithm families.

\subsection{Statistical interpretation}
No confidence interval or significance test is reported because the experiment evaluates every case in a deliberately selected public sample corpus. The executable identities are not a random sample from a defined population of ML-KEM inputs, and treating them as independent Bernoulli trials would manufacture a population model that the protocol does not justify. A statement such as ``100\% passed'' is descriptive of each declared corpus surface, not an estimate of an unknown universal success rate.

The three repetitions likewise do not create independent input diversity. They repeat the same deterministic inputs to detect instability. Counting them separately is appropriate for the raw execution record and the stability predicate, but not for claiming three times as much input diversity.

Performance inference is also excluded. Adapter elapsed time includes language runtime, process invocation, host scheduling, and harness overhead. No warm-up policy, calibrated clock analysis, power state control, or distributional benchmark design was predeclared. Using those diagnostic values for comparative performance would exceed the method.

\subsection{Independent reproduction agenda}
The source-built author run tests preparation completeness but does not close the independent-reproduction question. The next evidential step is a clean-environment rerun by a reviewer who did not author the harness. The reviewer should create a new work directory, verify every download, build all pinned releases, and compare terminal statuses with the preserved run. A different operating system or x86-64 host would add external-validity evidence; another arm64 macOS environment would still improve operational independence.

A stronger replication could implement an additional adapter independently from the disclosed adapter source while retaining the frozen case selection and equivalence rules. That would test whether the current adapters encode a shared interpretation mistake. A live ACVTS exercise, where available through the appropriate institutional process, would address a different question and must be reported separately from public sample replay.

Future corpus extensions should be predeclared and receive a new protocol revision, input manifest, and analysis plan rather than being appended after inspecting failures.

\section{Threats to Validity and Limitations}
\subsection{Construct validity}
Conformance here means byte or verdict equality on public sample vectors, not security, implementation quality, certification, or ecosystem-wide behaviour.

\subsection{Internal validity}
Adapter defects could misclassify results. The adapters are intentionally small and use different public execution paths for noble, liboqs, and Go. The runner preserves errors rather than converting them into failed validation verdicts or skips. The frozen control suite exercises the byte, verdict, and malformed-response paths, but independent review and execution remain necessary.

Validation-signal specificity is not uniform. Noble's allowlisted messages and Go's constructor errors distinguish the mapped rejection conditions, whereas liboqs exposes the generic \mbox{\texttt{OQS\_ERROR}} value at its public KEM boundary. Correct-length inputs, a pinned backend, driver-level allocation checks, and rejection behaviour documented for that backend constrain the interpretation, but they cannot prove that every possible backend \mbox{\texttt{-1}} denotes only key validation. A different internal failure returning the same public value could be misclassified as verdict \mbox{\texttt{false}}. The reported result is therefore conditional on the pinned backend path and its documented return contract.

All adapters compare against the same NIST files. A corpus error would therefore affect every provider. Pinning the corpus makes such an issue diagnosable but cannot rule it out. The study uses the public sample corpus, not vectors generated in a live ACVTS session.

\subsection{External validity}
Only one release of each implementation, one NIST corpus commit, one operating system, and one arm64 environment were measured. Results may differ on other architectures, compiler configurations, acceleration paths, or versions. No browser, WebAssembly, x86-64, or hardware implementation was tested.

\subsection{Coverage limitation}
Noble and liboqs executed the complete selected corpus, but Go executed only 70 of 240 identities per repetition. Its full key-generation output, decapsulation, decapsulation-key checking, and ML-KEM-512 surface remain unevaluated under this corpus and public-interface rule. No randomized fuzzing, fault injection, memory analysis, or side-channel tests were added beyond the pinned corpus.

\subsection{Independence limitation}
Different languages and project structures do not guarantee independent algorithm ancestry. The pinned liboqs release dispatches ML-KEM to vendored mlkem-native. Go's implementation entered the standard library from the filippo.io/mlkem768 lineage. Noble is separately maintained TypeScript source, but all providers share FIPS 203 and public test material. The experiment claims observational agreement across public packages, not genealogical independence.

\subsection{Oracle and corpus validity}
The NIST expected-result files are treated as the oracle for this study. A defect in those files or an incorrect prompt-to-expected join could make all providers appear to agree with an incorrect value. Exact commit and file digests make the selected oracle inspectable and stable, but they do not prove that the oracle is error-free. The use of public NIST material is a trust decision appropriate to the research question, not a formal derivation of expected outputs.

The selected corpus is also curated rather than population-representative. It covers three parameter sets and the available sample operations but does not sample uniformly from key, ciphertext, or randomness spaces. Conclusions cannot be extrapolated statistically to all inputs. The corpus is best understood as a named regression and conformance surface.

Join correctness is an additional threat. Test identifiers are scoped by the source structure, and an incorrect join could pair a prompt with the wrong expected result. The harness joins using group and case identifiers and preserves those identifiers in output, enabling an audit. Independent inspection of a sample from each operation and parameter set would further reduce this threat.

\subsection{Toolchain and environment validity}
The observed result depends on the build and execution environment. Compiler version, build configuration, JavaScript runtime, operating system, processor architecture, and enabled implementation paths can affect behaviour. The experiment records these dependencies but evaluates only one environment. In particular, it does not exercise x86-64 optimizations, WebAssembly, browser packaging, shared-library loading, or hardware acceleration paths.

The preparation script minimizes liboqs to the three ML-KEM variants needed by the protocol. That improves build focus but is not equivalent to every supported downstream build configuration. The noble release is built through its declared package process with lifecycle scripts disabled during installation. A different package manager, transpilation target, or bundler could produce a different integration surface even when algorithm source is unchanged.

Process boundaries also differ. Noble calls a JavaScript module, liboqs calls a linked C driver, and Go uses a source-built binary. A controller defect could interact differently with these paths. The study treats nonzero exit, panic, malformed output, unexpected provider signals, and invocation failure explicitly, but it does not measure resource leaks, long-run process stability, or concurrent execution.

\subsection{Reporting and artifact validity}
The author both developed the harness and interpreted the results, creating a risk of confirmation bias. Predeclared protocol fields, complete raw records, automated aggregate derivation, and explicit unsupported cases reduce but do not eliminate that risk. Independent reproduction and code review remain necessary before a mature reproducibility claim.

Machine-readable manifests protect against unnoticed artifact drift after generation, provided reviewers verify them. They do not prevent an author from selecting an inadequate corpus or writing an overbroad conclusion. Technical integrity controls and scholarly judgement are complementary.

Publication packaging creates a final risk: a correct research record can be transformed into an incomplete or contaminated release artifact. A file allowlist, provenance map, digest manifest, and prohibited-pattern scan are therefore required before release. These packaging controls do not strengthen the experimental result; they preserve the identity and public-only boundary of the reviewed artifact.

\section{Reproducibility}
The research artifact contains:

\begin{itemize}
\item immutable source and vector pins with SHA-256 digests;
\item a preparation script that downloads and verifies public sources and builds noble, minimal liboqs, and Go 1.26.4 from pinned source;
\item a provider-neutral runner and three public-interface adapters;
\item the frozen capability matrix, error taxonomy, namespaces, equivalence rules, timing policy, and audit predicates;
\item all 2,160 base records and 150 isolated Go key-generation projection records;
\item the frozen byte-oracle and verdict-oracle mutation specifications and their respective 240-case results;
\item the frozen malformed-adapter-response specification and its one-error, zero-false-verdict result;
\item summary, environment, and input manifests; and
\item a result-artifact manifest containing SHA-256 digests and byte counts.
\end{itemize}
Preparation requires network access. After preparation, measurement runs offline. A rerun should verify downloads, execute the runner into a new output directory, confirm every audit predicate, and compare scientific fields rather than elapsed times or generation timestamps. An independent rerun on a second architecture would materially strengthen the external-validity evidence.

For a strict comparison, a reviewer should reduce each raw record to \texttt{\mbox{(namespace},\allowbreak{}\mbox{ provider},\allowbreak{}\mbox{ repetition},\allowbreak{}\mbox{ operation},\allowbreak{}\mbox{ parameterSet},\allowbreak{}\mbox{ tgId},\allowbreak{}\mbox{ tcId},\allowbreak{}\mbox{ status},\allowbreak{}\mbox{ reason},\allowbreak{}\mbox{ observedVerdict},\allowbreak{}\mbox{ mismatchFields)}}, sort deterministically, and compare the resulting collections. Expected differences such as run timestamps, absolute paths, and diagnostic durations should be excluded. Any missing tuple, additional tuple, changed terminal status, verdict, mismatch field, or unsupported reason is a reproduction discrepancy.

Artifact verification proceeds in two layers. The result-artifact manifest binds files generated by the author run. The candidate release manifest binds every proposed public file, including the manuscript, product-neutral source, result records, notices, and vector figures. The second manifest is not a substitute for the first: one establishes internal consistency of a run, while the other establishes exact composition and provenance of a proposed release.

Standalone figures repeat only values already present in the raw results and numbered tables; they introduce no independent measurement.

A successful rerun should be recorded as a separate evidence event rather than overwriting the author run. It should name the reviewer or responsible organization as appropriate, execution date, environment, source and vector digest verification outcome, raw-result digest, comparison procedure, and every discrepancy. Preserving both events supports an audit trail and prevents a later successful run from erasing an earlier failure.

The artifact is archived at Zenodo under DOI \texttt{\mbox{10.5281}/\allowbreak{}\mbox{zenodo.21910571}} and released under the MIT License. Independent reproduction has not yet occurred and is recorded separately when it does.

\section{Conclusion}
Three pinned public ML-KEM packages reproduced every expected byte or validation verdict on their declared executable surfaces of a pinned NIST ACVP sample corpus. Across three repetitions, all 1,650 executable evaluations passed, all 510 unsupported records matched Go's frozen capability boundary, and no base failure, adapter error, pairwise-overlap disagreement, or status instability occurred. A separate Go projection matched 150 encapsulation keys without being counted as full key generation. Frozen byte, verdict, and malformed-response controls each produced the exact predeclared failure or error and no false verdict. The central contribution is the evidence protocol around these results: exact pins, provider-specific capability contracts, explicit validation-error taxonomy, case-preserving records, namespace isolation, and formal audit predicates make the measurement independently auditable and falsifiable.

The finding is useful but bounded. It demonstrates author-run repeatability for the selected outcomes on the reported host and exposes a public-interface testability gap; it does not demonstrate independent external reproduction. It does not certify a provider, prove security, test side channels, validate integration, or authorize production claims. Future work should seek independently operated reruns on additional architectures and separately frozen specification-divergence challenges.

\section{Declarations}
\textbf{Acknowledgements and research context:} BEE (the Progressive Quantum-Native Intelligence Engine), by HEOSSI, supported research organization, source review, evidence orchestration, and artifact verification under author supervision. QNSI (Quantum-Native Security Infrastructure) provided the governed research and evidence-control environment. Neither BEE nor QNSI was an evaluated implementation, and no private product code, customer material, production credentials, or operational system entered the experiment.

\textbf{Funding:} This study received no external funding.

\textbf{Competing interests:} The author is associated with HEOSSI (Pte.) Ltd., which publishes related cryptographic-conformance evidence; that evidence was not an input or oracle here.

\textbf{Data and code availability:} The product-neutral source, frozen specifications, and result records are archived at DOI \texttt{\mbox{10.5281}/\allowbreak{}\mbox{zenodo.21910571}} under the MIT License.

\textbf{Ethics:} The study used no human participants, personal data, customer data, credentials, or production systems.

\textbf{Author contributions:} Christopher M. Frost conceived and supervised the study, defined its public-only boundary, verified its claims and evidence, and is responsible for the manuscript and release decision.

\textbf{Use of generative AI.} Generative AI tools provided general assistance with research organization, source review, and manuscript drafting and editing.

\balance
\fontsize{7.8}{8.9}\selectfont


\begin{thebibliography}{99}
\setlength{\itemsep}{0pt}
\setlength{\parskip}{0pt}
\bibitem{ref1} National Institute of Standards and Technology, ``Module-Lattice-Based Key-Encapsulation Mechanism Standard,'' FIPS 203, Aug. 2024, \href{https://doi.org/10.6028/NIST.FIPS.203}{doi: 10.6028/NIST.FIPS.203}.\par
\bibitem{ref2} National Institute of Standards and Technology, ``Cryptographic Algorithm Validation Program,'' 2026. [Online]. Available: \url{https://csrc.nist.gov/projects/cryptographic-algorithm-validation-program}. Accessed: Aug. 11, 2026.\par
\bibitem{ref3} National Institute of Standards and Technology, ``ACVP-Server,'' commit \texttt{\seqsplit{a7f283cdc87d2d6dd93c1bac59e5622c5f9f8324}}, July 31, 2026. [Online]. Available: \url{https://github.com/usnistgov/ACVP-Server/commit/a7f283cdc87d2d6dd93c1bac59e5622c5f9f8324}.\par
\bibitem{ref4} National Institute of Standards and Technology, ``ACVP ML-KEM JSON Specification,'' 2026. [Online]. Available: \url{https://pages.nist.gov/ACVP/draft-celi-acvp-ml-kem.html}. Accessed: Aug. 11, 2026.\par
\bibitem{ref5} P. Miller et al., ``noble-post-quantum,'' version 0.7.0, Aug. 2026. [Online]. Available: \url{https://github.com/paulmillr/noble-post-quantum/releases/tag/0.7.0}.\par
\bibitem{ref6} Open Quantum Safe, ``liboqs,'' version 0.16.0, July 2026. [Online]. Available: \url{https://github.com/open-quantum-safe/liboqs/releases/tag/0.16.0}.\par
\bibitem{ref7} D. Stebila and M. Mosca, ``Post-quantum key exchange for the Internet and the Open Quantum Safe project,'' in Selected Areas in Cryptography - SAC 2016, LNCS 10532, pp. 14-37, 2017, \href{https://doi.org/10.1007/978-3-319-69453-5_2}{doi: 10.1007/978-3-319-69453-5\_2}.\par
\bibitem{ref8} R. Avanzi et al., ``CRYSTALS-Kyber algorithm specifications and supporting documentation,'' version 3.02, Aug. 2021. [Online]. Available: \url{https://pq-crystals.org/kyber/data/kyber-specification-round3-20210804.pdf}.\par
\bibitem{ref9} J. Bos et al., ``CRYSTALS-Kyber: A CCA-secure module-lattice-based KEM,'' in 2018 IEEE European Symposium on Security and Privacy, pp. 353-367, 2018, \href{https://doi.org/10.1109/EuroSP.2018.00032}{doi: 10.1109/EuroSP.2018.00032}.\par
\bibitem{ref10} National Institute of Standards and Technology, ``Automated Cryptographic Validation Protocol Documentation,'' 2026. [Online]. Available: \url{https://pages.nist.gov/ACVP/}. Accessed: Aug. 11, 2026.\par
\bibitem{ref11} Association for Computing Machinery, ``Artifact Review and Badging, Version 1.1,'' Aug. 2020. [Online]. Available: \url{https://www.acm.org/publications/policies/artifact-review-and-badging-current}. Accessed: Aug. 11, 2026.\par
\bibitem{ref12} G. Alagic et al., ``Status Report on the Third Round of the NIST Post-Quantum Cryptography Standardization Process,'' NIST IR 8413-upd1, Sept. 2022, \href{https://doi.org/10.6028/NIST.IR.8413-upd1}{doi: 10.6028/NIST.IR.8413-upd1}.\par
\bibitem{ref13} W. M. McKeeman, ``Differential testing for software,'' Digital Technical Journal, vol. 10, no. 1, pp. 100-107, 1998. [Online]. Available: \url{https://dblp.org/rec/journals/dtj/McKeeman98.html}.\par
\bibitem{ref14} The C2SP Project, ``Project Wycheproof,'' 2026. [Online]. Available: \url{https://github.com/C2SP/wycheproof}. Accessed: Aug. 13, 2026.\par
\bibitem{ref15} G. Alagic et al., ``Recommendations for Key-Encapsulation Mechanisms,'' NIST SP 800-227, Sept. 2025, \href{https://doi.org/10.6028/NIST.SP.800-227}{doi: 10.6028/NIST.SP.800-227}.\par
\bibitem{ref16} National Institute of Standards and Technology, ``Security Requirements for Cryptographic Modules,'' FIPS 140-3, Mar. 2019, \href{https://doi.org/10.6028/NIST.FIPS.140-3}{doi: 10.6028/NIST.FIPS.140-3}.\par
\bibitem{ref17} National Institute of Standards and Technology, ``Secure Hash Standard,'' FIPS 180-4, Aug. 2015, \href{https://doi.org/10.6028/NIST.FIPS.180-4}{doi: 10.6028/NIST.FIPS.180-4}.\par
\bibitem{ref18} National Academies of Sciences, Engineering, and Medicine, Reproducibility and Replicability in Science. Washington, DC, USA: National Academies Press, 2019, \href{https://doi.org/10.17226/25303}{doi: 10.17226/25303}.\par
\bibitem{ref19} M. D. Wilkinson et al., ``The FAIR Guiding Principles for scientific data management and stewardship,'' Scientific Data, vol. 3, art. 160018, 2016, \href{https://doi.org/10.1038/sdata.2016.18}{doi: 10.1038/sdata.2016.18}.\par
\bibitem{ref20} A. M. Smith, D. S. Katz, and K. E. Niemeyer, ``Software citation principles,'' PeerJ Computer Science, vol. 2, e86, 2016, \href{https://doi.org/10.7717/peerj-cs.86}{doi: 10.7717/peerj-cs.86}.\par
\bibitem{ref21} T. Bray, ``The JavaScript Object Notation (JSON) Data Interchange Format,'' RFC 8259, Dec. 2017, \href{https://doi.org/10.17487/RFC8259}{doi: 10.17487/RFC8259}.\par
\bibitem{ref22} R. Barnes, K. Bhargavan, B. Lipp, and C. Wood, ``Hybrid Public Key Encryption,'' RFC 9180, Feb. 2022, \href{https://doi.org/10.17487/RFC9180}{doi: 10.17487/RFC9180}.\par
\bibitem{ref23} E. Barker and J. Kelsey, ``Recommendation for Random Number Generation Using Deterministic Random Bit Generators,'' NIST SP 800-90A Rev. 1, June 2015, \href{https://doi.org/10.6028/NIST.SP.800-90Ar1}{doi: 10.6028/NIST.SP.800-90Ar1}.\par
\bibitem{ref24} The Go Project, ``Package crypto/mlkem,'' Go 1.26.4 standard-library documentation, 2026. [Online]. Available: \url{https://pkg.go.dev/crypto/mlkem@go1.26.4}. Accessed: Aug. 13, 2026.\par
\bibitem{ref25} National Institute of Standards and Technology, ``Accessing the Automated Cryptographic Validation Testing System,'' 2026. [Online]. Available: \url{https://csrc.nist.gov/projects/cryptographic-algorithm-validation-program/how-to-access-acvts}. Accessed: Aug. 13, 2026.\par
\bibitem{ref26} G. Vranken, ``Cryptofuzz: Differential cryptography fuzzing,'' 2019. [Online]. Available: \url{https://github.com/MozillaSecurity/cryptofuzz} (maintained fork). Accessed: Aug. 13, 2026.\par
\bibitem{ref27} O. O. Egbuagha, ``A differential fuzzing framework for ML-KEM and ML-DSA implementations in cryptographic libraries,'' Master's thesis, University of Oulu, Oulu, Finland, 2026. [Online]. Available: \url{https://oulurepo.oulu.fi/handle/10024/63293}.\par
\bibitem{ref28} J.-P. Aumasson and Y. Romailler, ``Automated testing of crypto software using differential fuzzing,'' Black Hat USA, Las Vegas, NV, USA, July 2017. [Online]. Available: \url{https://github.com/kudelskisecurity/cdf}. Accessed: Aug. 13, 2026.\par
\end{thebibliography}
\end{document}